Gate-tunable spin-charge conversion and a role of spin-orbit interaction in graphene

S. Dushenko[1,2], H. Ago[3], K. Kawahara[3], T. Tsuda[4], S. Kuwabata[4], T. Takenobu[5], T. Shinjo[2], Y. Ando[2], and M. Shiraishi[2,*]

1. Graduate School of Engineering Science, Osaka University, Toyonaka 560-8531, Japan

2. Department of Electronic Science and Engineering, Kyoto University, Kyoto 615-8510, Japan

3. Institute for Material Chemistry and Engineering, Kyushu University, Fukuoka 816-8508, Japan

4. Graduate School of Engineering, Osaka University, Suita 565-0871, Japan

5. School of Advanced Science and Engineering, Waseda University, Tokyo 169-8555, Japan

* Corresponding author: Masashi Shiraishi (mshiraishi@kuee.kyoto-u.ac.jp)

**The small spin-orbit interaction of carbon atoms in graphene promises a long spin diffusion length and potential to create a spin field-effect transistor. However, for this reason, graphene was largely overlooked as a possible spin-charge conversion material. We report electric gate tuning of the spin-charge conversion voltage signal in a single-layer graphene. Using spin pumping from yttrium iron garnet ferrimagnetic insulator and ionic liquid top gate we determined that the inverse spin Hall effect is the dominant spin-charge conversion mechanism in a single-layer graphene. From the gate dependence of the electromotive force we showed dominance of the intrinsic over Rashba spin-orbit interaction: a long-standing question in graphene research.**

Spintronics promises salvation from the scaling limitations of electronics. However, to create spin-logic devices, long spin transport and control of the spin current must be achieved. In this light, single-layer graphene (SLG) [1]—a two-dimensional sheet of carbon atoms bound together by sp$^2$ hybridized orbitals in a honeycomb lattice—has received tremendous attention as a potential material for spintronic devices. Graphene has a long spin relaxation time and spin diffusion length [2,3], and, due to its two-dimensional nature, its electronic properties can be easily modulated by gate voltage [4]. Another important requirement to create spintronic devices is the ability to convert spin current into charge current and vice versa. The spins of the carriers are coupled to the angular momentum of the lattice atoms through the spin-orbit interaction (SOI). However, SOI is not restricted to the lattice atoms and fundamentally originates from the coupling between the electric field and the moving spin, and enables the conversion of the pure spin current into the charge current (inverse spin Hall effect [5,6], inverse Rashba-Edelstein effect [7,8]) and reciprocally (spin Hall effect [9,10] Rashba-Edelstein effect [11–13]). In this recently-emerged experimental spin-charge conversion branch of spintronics, graphene was largely overlooked due to the very small intrinsic SOI of carbon atoms, with only a few experimental studies performed [14,15]. Despite many theoretical studies, there is still no agreement in the literature today on the strength and type of the spin-charge conversion and underlying SOI mechanism in graphene (dependent or independent on electric field). Our aim in this study was to experimentally distinguish between SOI mechanisms in SLG. By measuring the gate voltage dependence of the spin-charge conversion in SLG on top of ferrimagnetic insulator yttrium iron garnet (YIG), we determined the dominant type of the spin-charge conversion mechanism: a long-standing question in graphene research.

We used spin pumping [16,17] to generate a pure spin current in the SLG. In spin pumping, precession of the magnetization in the ferromagnetic layer drives pure spin current through the ferromagnetic interface. Spin current transfers the spin-angular momentum from the ferromagnetic layer to the adjacent layer, thus leading to enhanced relaxation of the magnetization precession under ferromagnetic resonance (FMR) conditions. Spin pumping was shown to overcome the conductance mismatch problem [18], and has recently been widely used for spin injection in semiconductors [19], conjugated polymers [20] and SLG [21]. As a ferrimagnetic insulator, YIG has great potential as a spin source due to its low magnetic damping and band gap of 2.7 eV at room temperature. The large band gap prevents spurious voltage signals from the ferrimagnetic layer, which are present in the case of common metallic ferromagnets [22]. Spin injection into graphene using spin pumping recently received attention both from theoretical [23] and experimental [14,21] studies. In our study—using spin injection from the YIG substrate via spin pumping—we studied SOI in SLG and achieved spin-charge conversion tuning using gate voltage application.

A schematic layout of the experiment is shown in Fig. 1. Precession of the magnetic moment in the YIG layer under FMR conditions led to the flow of the angular momentum into the adjacent SLG [Fig. 1(a)]. In the SLG, due to the spin-charge conversion, the spin current was converted into a charge current. The generated charge current was detected as a voltage at the Ti/Au contact pads on the opposite sides of the sample [Fig. 1(c)]. The carrier density and carrier type were controlled via an ionic liquid electric gate on top of the SLG [Fig. 1(b)] (see Supplemental material, which includes references [24–46], for the detailed experimental methods).

We now proceed to the experimental results. We used high-quality SLG grown by chemical vapor deposition using an atomically flat epitaxial Cu(111) thin film [47]. Figure 2 shows the typical Raman spectra of the SLG after transfer on the $SiO_2$ (red line) and YIG (blue line) substrates. Orange and black lines shows Raman spectrum of YIG before single layer graphene transfer and difference Raman spectrum YIG/SLG–YIG, respectively. The $SiO_2$/SLG and YIG/SLG–YIG Raman spectra showed sharp G and 2D-bands with a high relative intensity ($I_{2D}/I_G$) > 2. In addition, D-band at 1340 cm$^{-1}$, which originates in the structural defects and edges, was negligible. These results confirmed the growth and successful transfer of high-quality SLG [48]. Residual peak from YIG substrate at 2520 cm$^{-1}$ is present in the difference spectrum due to change in the intensity of the YIG bands between YIG and YIG/SLG samples.

Figures 3(a) and 3(b) show the FMR spectrum with the external magnetic field applied in the plane of the sample surface. Microwave field $h_{rf}$ excited resonances in the FMR spectrum through coupling to the magnetostatic spin-waves in the YIG layer. The peak of the absorbance spectrum at $H_{FMR}$ corresponds to the fundamental wave mode with $k = 0$, which is equivalent to the uniform FMR mode in the ferromagnetic layers. Slight asymmetry of the absorbance spectrum in the in-plane external magnetic field configuration was caused by magnetostatic surface waves at $H < H_{FMR}$ and magnetostatic backward volume waves at $H > H_{FMR}$ [49,50]. From the FMR spectrum, the saturation magnetization of YIG was $\mu_0 M = 0.169$ T at room temperature (using g-factor $g = 2.046$ [51]), which agrees well with the values reported for YIG films in the literature [50,51]. The saturation magnetization at the temperature of spin pumping measurements ($T$ = 200 K) was $\mu_0 M$ = 0.204 T, which was in accordance with a temperature dependence of $\mu_0 M$ from Neel's two-sublattice model [52]. Though in metallic ferromagnet systems the bias voltage can influence the spin pumping efficiency through the control of the depletion region and Schottky barrier height, spin pumping from the ferrimagnetic insulator with large 2.7 eV bang gap is free of this problem. Additionally, we measured linear *I-V* curve at room and low temperatures, ruling out presence of any energy barriers at the Au/Ti/SLG interface. Thus, spin current injection and FMR spectrum are expected to be independent of gate voltage. Detected FMR absorbance spectra [Figs. 3(a) and 3(b)] were identical for 0.5 V (dashed red line) and 1.0 V (black line). The drain current $I_D$ showed strong $V_G$ dependence, confirming the carrier density modulation and switching of the carrier type from holes to electrons [see Fig. 4(a)]. In contrast, both the full width at half maximum, $\Delta H_{FWHM}$ [Fig. 3(e)], and the amplitude, $I_{Sym}$ [Fig. 3(f)], obtained from the fitting of the FMR spectrum were independent of $V_G$ and lied in the range: $\mu_0 \Delta H_{FWHM}^{0°}$ = 0.545 mT ($^{+0.5}_{-0.6}$ %), $\mu_0 \Delta H_{FWHM}^{180°}$ = 0.540 mT ($^{+1.1}_{-0.9}$ %), $I_{Sym}^{0°}$ = 436 arb. unit ($^{+0.5}_{-0.4}$ %), and $I_{Sym}^{180°}$ = 398 arb. unit ($^{+1.5}_{-1.2}$ %) (see section E and G of Supplemental material for the fitting function and detailed discussion on FMR spectra). The small difference in the FMR at the opposite directions of external magnetic field was caused by differences in the spin wave resonances due to slightly different sample position at $\theta_H$ = 0° and $\theta_H$ = 180°.

Next, we discuss electrical response of the YIG/SLG under gate voltage application. It was shown that spin current generated through the spin Hall effect can be used to switch magnetization in spin-torque devices [53]. However, until the present, sign of the generated spin and charge current can only be changed by reversal of the magnetic field or source current. Fig. 4(a) shows the dependence of the drain current $I_D$ between Ti/Au terminals under sweep of the gate voltage $V_G$ for our YIG/SLG device. The minimum of the $I_D$ curve represents the position of the Dirac point in the $V_G$ sweep. Figures 4(b) and 4(c) show the detected electromotive force under the FMR conditions. The generated spin current, i.e., transferred angular momentum from YIG to SLG, is proportional to the area swept by the YIG magnetization during the precession. Thus, spin current has maximum at the FMR field $H_{FMR}$: at which microwave power absorption and precession angle are the largest. As a result, electromotive force generated through the spin-charge conversion has a symmetrical shape with respect to the $H_{FMR}$. The measured voltage signals were deconvolved into symmetrical $V_{\text{Sym}}$ and $V_{\text{Asym}}$ contributions using the Lorentzian functions [54]. The small asymmetric contribution $V_{\text{Asym}}$ to the voltage signal may have arisen from spurious effects, which include thermal effects (see section E of Supplemental material for details). At $V_G$ = 0.5 V [red line in Figs. 4(b) and 4 (c)], the Fermi level in SLG was tuned below the Dirac point, where holes acted as a dominant carrier. When $V_G$ was tuned to 1.0 V [blue line in Figs. 4(b) and 4(c)], the Fermi level crossed the Dirac point, and the carrier type was changed from holes to electrons. The change in the carrier type was accompanied by switching of the voltage signal sign for both $\theta_H = 0°$ and $\theta_H = 180°$ directions of the external magnetic field [Figs. 4(b) and 4(c), respectively]. Thus, a gate voltage can be used to switch sign of the charge current generated through SOI.

In Fig. 5(a), the gate voltage dependence of $V_{\text{Sym}}$ for two external magnetic field **H** directions $\theta_H = 0°$ (black filled circles) and $\theta_H = 180°$ (red filled circles) is shown. Using the resistance determined from the $I_D$-$V_G$ measurements and the detected voltage $V_{\text{Sym}}$, we extracted the charge current $I$ generated in the YIG/SLG system. Contributions to the spin-charge conversion current from holes and electrons have opposite polarities. Hence, the charge current generated in the SLG was described by the following relation: $I = I_S \frac{n_h - n_e}{n_h + n_e}$, where $I_S = \lim_{V_G \to -\infty} I(V_G)$ is the charge current when the Fermi level is tuned far below Dirac point. Under the assumption that the carrier type is switched from holes to electrons at the Dirac point, the dependence of the carrier type $n_{\text{Type}}$ on the gate voltage is described by $n_{\text{Type}} = -\text{sgn}(V_G - V_{\text{DP}})$, where $n_{\text{Type}} = 1$ corresponds to the hole carrier type and $n_{\text{Type}} = -1$ corresponds to the electron carrier type. However, due to fluctuations of the Fermi level in devices, both electrons and holes are present near the Dirac point (phenomenon known as electron-hole puddles). To take into account presence of the electron-hole puddles near the Dirac point, we assumed the following distribution of the normalized electron and hole carrier densities near the Dirac point: $\frac{n_h}{n_h + n_e} = \frac{1}{e^{(V_G - V_{\text{DP}})/a} + 1}$ and $\frac{n_e}{n_h + n_e} = \frac{e^{(V_G - V_{\text{DP}})/a}}{e^{(V_G - V_{\text{DP}})/a} + 1}$. The generated charge current consisted of two components: the contribution from the spin pumping $I_S$, which reverses its sign at the reversal of the external magnetic field from $\theta_H = 0°$ to $\theta_H = 180°$, and thermal component $I_{\text{Th}}$ from the heating effects, which was independent of the magnetic field direction. The parameters ($V_{\text{DP}}$, $a$), dependent on ionic-gating, can vary to some extent in different sweeps of the gate voltage (for $\theta_H = 0°$ and $\theta_H = 180°$) due to structural changes in the electric double layer of the ionic liquid. However, the charge current parameters, $I_S$ and $I_{Th}$, are independent of the ionic-gate sweep, because they are determined by the spin pumping efficiency, which is independent of $V_G$. From the fitting of the experimental data we obtain values $I_S = 1.05$ nA, $I_{Th} = 0.32$ nA, $V_{DP}^{\theta_H=0°} = 0.9\,V$, $a^{\theta_H=0°} = 0.25\,V$, $V_{DP}^{\theta_H=180°} = 0.6\,V$, and $a^{\theta_H=180°} = 0.19\,V$. Figures 5(a) and 5(b) show experimental data fitting by model with the switching of the single carrier type at the Dirac point (dashed lines) and model with electron-hole puddle transition (solid lines).

We now consider the origin of the spin-charge conversion mechanism in SLG in our system. The intrinsic SOI of graphene is responsible for the inverse spin Hall effect, and is present even when reflection symmetry with respect to graphene plane is not broken. The Rashba SOI, which arises in the presence of the electric field $E$ perpendicular to the surface of the graphene layer and broken reflection symmetry, is responsible for the spin-charge conversion via the inverse Rashba-Edelstein effect. There is no agreement in the literature whether the intrinsic-like SOI or the Rashba-like SOI (in the range of commonly applied electric fields $10^8$-$10^9$ V/m) is dominant in pristine graphene, and estimations for both of them span over a wide range of $10^{-3}$-$10^1$ meV, 4 orders of magnitude [55–60]. The issue is non-trivial for future graphene-based applications: the intrinsic-like SOI leads to the opening of a topological gap at the K point, giving an effective mass to carriers and turning graphene into a quantum spin Hall insulator [58]. In contrast, the Rashba-like SOI interaction prevents the gap from opening [56,58] but can be used for spin control in spin transport experiments [61]. The Rashba SOI parameter $\alpha_R$ depends linearly on the electric field $E$ induced by the gate voltage: $\alpha_R \propto Ee\xi$, where $\xi$ is the strength of the atomic SOI [55,56,62]. Thus, amplitude of the charge current generated by the inverse Rashba-Edelstein effect linearly depends on the applied gate voltage. In contrast, we observed a constant amplitude of the current $I_S = 1.05$ nA, which was independent of the gate voltage $V_G$ far from the Dirac point, and switched polarity in the vicinity of $V_{DP}$ [Fig. 5(b)]. It indicates that contribution of the Rashba SOI in graphene is negligibly small, and rules out the inverse Rashba-Edelstein effect as a spin-charge conversion mechanism. Thus, the inverse spin Hall effect is the most probable mechanism for spin-charge conversion in the YIG/SLG system. After the submission of the manuscript the authors became aware of another work that addresses spin-charge conversion in the YIG/SLG [63]. In the spin-charge conversion part of their study Mendes et al. [63] carried out the same spin pumping-induced spin-charge conversion measurements as previously reported by Ohshima et al. [14]. While the same experimental results were observed in both studies, the given interpretations were strikingly different: in one study the inverse spin Hall effect was suggested as the origin of the observed electromotive force [14], in the other study—the inverse Rashba-Edelstein effect [63]. In our study, in contrast to the previous works, we applied out of plane electric field to tune SOI and probe the spin-charge conversion mechanism. Independence of the spin-charge conversion current from the applied top gate voltage for the first time showed unequivocal experimental evidence that the inverse spin Hall effect, and not the inverse Rashba-Edelstein effect, is the dominant spin-charge conversion mechanism in the SLG.

Finally, we consider the possible enhancement of SOI in the studied SLG due to the extrinsic effects. Recent studies on graphene heavily decorated with adatoms reported that the skew scattering can lead to the large intrinsic-like SOI and inverse spin Hall effect [15,64]. However, this mechanism is not applicable in our case, since large density of adatoms clusters was absent in the studied SLG (see section H of Supplemental material for the details). Another mechanism that can induce large enhancement of the SOI is the proximity effect, similar to the one reported in graphene on top of the $WS_2$ substrate [57,65]. However, the proximity-induced SOI in graphene systems is quickly suppressed with the increased distance between graphene layer and the substrate [65–67]. In [68]—where the magnetic proximity effect was observed experimentally via the anomalous Hall effect—the YIG substrate was atomically flat with roughness on the terraces ~0.06 nm. In our samples typical root mean square roughness of YIG substrates lay within range 0.3-0.5 nm. Thus, magnetic proximity effect was most probably suppressed in the studied system due to the large interlayer distance between SLG and YIG substrate.

In summary, we experimentally clarified that the intrinsic-like SOI is dominant over the Rashba-like SOI in the SLG in the range of commonly applied electric fields ($10^8$-$10^9$ V/m). It was shown that the inverse spin Hall effect is the dominant spin-charge conversion mechanism in SLG, and that polarity of the spin-charge conversion current can be switched by gate voltage application.

This work was supported by MEXT (Innovative Area "Nano Spin Conversion Science" KAKENHI No. 26103003), Grant-in-Aid for Challenging Exploratory Research (No.25630148) and Research Grant from Izumi Science and Technology Foundation. H.A. and K.K. acknowledge PRESTO-JST for the support.

**Figures**

FIG. 1

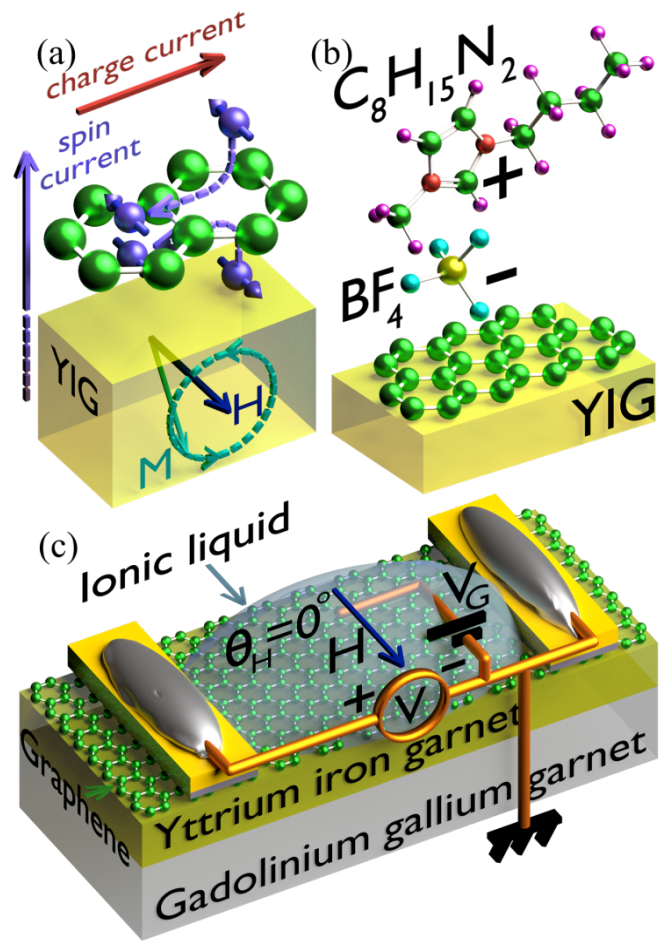

FIG. 2

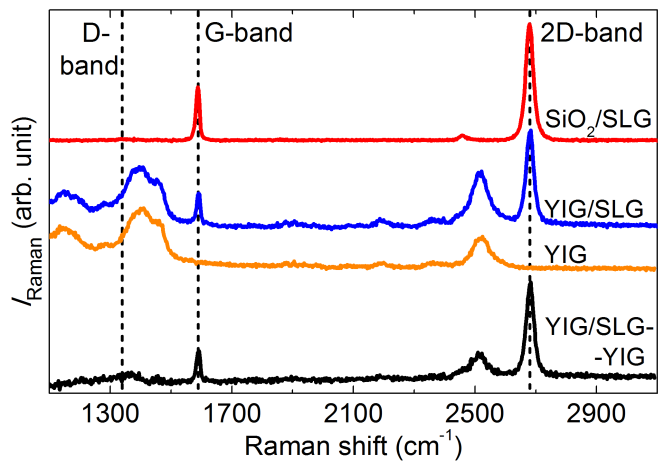

FIG. 3

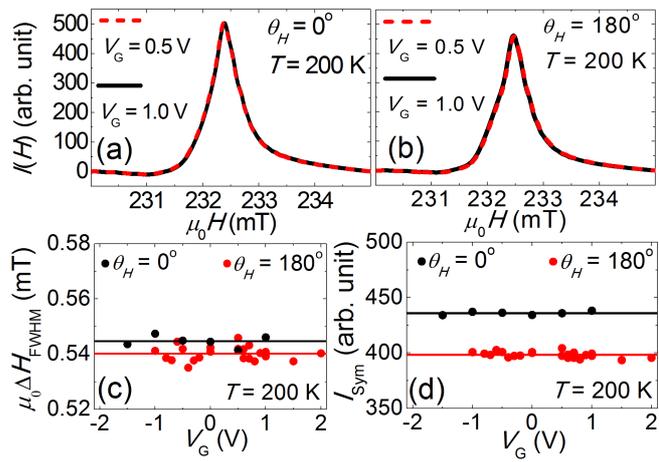

FIG. 4

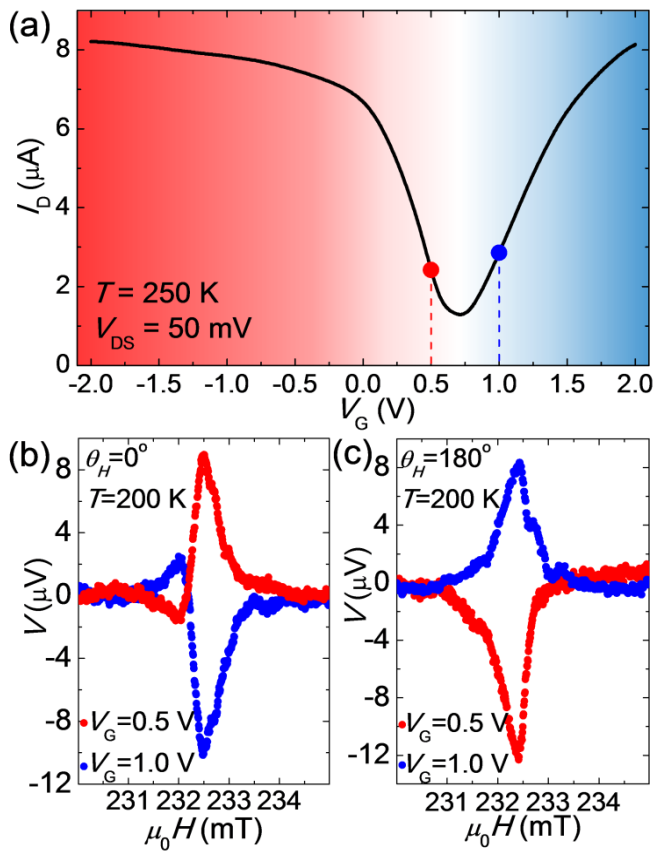

FIG. 5

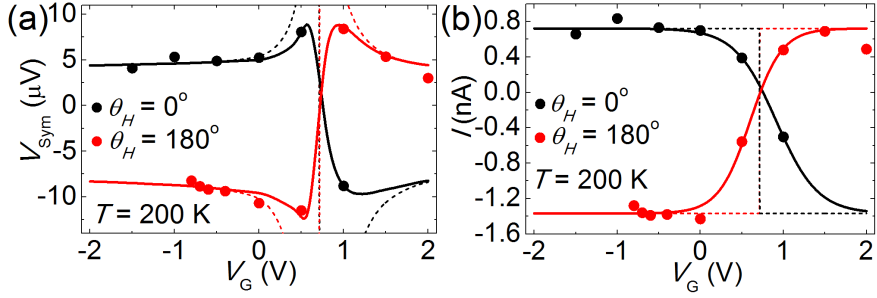

**Figure captions**

FIG. 1 (color online). (a) Layout of the spin-charge conversion experiment in the YIG/SLG sample. Under FMR pure spin current was transferred through the YIG/SLG interface. The pure spin current in the SLG was converted into an in-plane charge current. The generated voltage was detected from the Ti/Au contact pads. (c) Schematic view of an electric gate using ionic liquid placed on top of the SLG.

FIG. 2 (color online). The Raman spectra of the YIG substrate without single layer graphene (orange line), YIG/SLG (black line), difference Raman spectrum YIG/SLG–YIG (black line), and reference Si/SiO$_2$/SLG (red line) samples. Both SLG were grown in the same batch and were afterwards transferred to the designated substrates.

FIG. 3 (color online). FMR spectra under gate voltage application $V_G$ = 0.5 V (red dashed line) and $V_G$ = 1.0 V (black line) for $\theta_H$ = 0° (a) and $\theta_H$ = 180° (b). The FMR full width at half maximum $\Delta H_{FWHM}$ (c) and the FMR amplitude $I_{Sym}$ (d) dependence on the gate voltage $V_G$. Lines and solid circles represent the fitting by the constant and experimental data correspondingly.

FIG. 4 (color online). (a) $I_D$-$V_G$ measurement at $T$ = 250 K. In the red area, the carrier type is holes; in the blue area, the carrier type is electrons. Generated voltage under the FMR and application of the gate voltage $V_G$ = 0.5 V (red lines) and $V_G$ = 1.0 V (blue lines) for the direction of the external magnetic field **H** $\theta_H$ = 0° (b) and $\theta_H$ = 180° (c).

FIG. 5 (color online). $V_{Sym}$ (a) and spin-charge conversion current (b) dependence on the gate voltage $V_G$ for the direction of the external magnetic field **H** $\theta_H$ = 0° (black filled circles) and $\theta_H$ = 180° (red filled circles). Lines show the fitting that takes into account the presence of the electron-hole puddles near the Dirac point (solid lines) and the fitting using a simple one carrier type model (dashed lines).